\documentclass[a4paper,11pt]{article}
\usepackage{pos}

\title{Hyperon Physics at BESIII}

\author[a,b]{Hai-Bo Li}

\author*[a,c]{Hong-Fei Shen}
\onbehalf{on behalf of the BESIII collaboration}

\affiliation[a]{Institute of High Energy Physics,\\ Beijing 100049, People’s Republic of China}

\affiliation[b]{University of Chinese Academy of Sciences\\Beijing 100049, People's Republic of China}
\affiliation[c]{China Center of Advanced Science and Technology,\\ Beijing 100190, People’s Republic of China}

\emailAdd{shenhongfei@ihep.ac.cn}

\abstract{
    This proceeding presents recent advances in the study of hyperon physics using data from the BESIII experiment. 
    The BESIII detector has been in full operation at the BEPCII collider since 2008, providing excellent resolution, particle identification (PID), 
    and large coverage for both neutral and charged particles.
    Leveraging the outstanding capability of the detector, the BESIII experiment has collected 10 billion $J/\psi$ and 2.7 billion $\psi(3686)$ events.
    In recent years, BESIII has conducted a series of analyses focusing on hyperon physics, utilizing the pair production of quantum-entangled hyperon–antihyperon pairs from these charmonium decays.
    The transverse polarizations of the $\Lambda$, $\Sigma^{+,0}$, and $\Xi^{-,0}$ hyperons have been observed in $J/\psi$ and $\psi(3686)$ decays,
    providing excellent opportunities to search for the CP violation in hyperon decays. 
    Additionally, BESIII investigates weak radiative hyperon decays, semi-leptonic hyperon decays, and hyperon-nucleon interactions.
    }

    \FullConference{The 21st International Conference on Hadron Spectroscopy and Structure (HADRON2025)\\
 27 - 31 March, 2025\\
Osaka University, Japan\\}

\begin{document}
\maketitle

\section{Introduction}

Investigating the violation of the combined charge-conjugation and parity symmetry (CP violation) remains a fundamental pursuit in particle physics, 
as it is the key condition to explain the predominance of matter over antimatter in the universe. 
CP violation has been robustly observed in the kaon sector~\cite{Christenson:1964fg}, as well as in B~\cite{BaBar:2001pki, Belle:2001zzw} and D meson systems~\cite{LHCb:2019hro}.
More recently, the first discovery for CP violation in the baryon sector has been found in $\Lambda_b$ decays~\cite{LHCb:2025ray}. 
However, CP violation remains unobserved in hyperon decays, leaving open important questions in this area. 

Thanks to its ability to produce large samples of quantum-entangled hyperon–antihyperon pairs, 
the BESIII experiment offers a unique platform to explore potential CP violation in hyperon systems~\cite{Li:2016tlt,Donoghue:1986hh}. 
Utilizing the unprecedented datasets of $10^{10}$ $J/\psi$~\cite{BESIII:2021cxx} and $2.7\times10^9$ $\psi(3686)$~\cite{BESIII:2024lks} events, 
BESIII has performed a comprehensive series of analyses on the correlated production and decay of hyperons and antihyperons. 
These studies have led to precise determinations of hyperon properties and decay parameters, 
significantly advancing our understanding of CP symmetry and contributing to the broader knowledge in particle physics.
BESIII has also conducted extensive studies about hyperon radiative and semi-leptonic decays, 
which provide crucial insights for enhancing our understanding of non-perturbative QCD dynamics and for testing lepton universality.

The studies of hyperon-nucleon interactions are crucial for determining the equation of state of nuclear matter at supersaturation densities 
and for understanding the so-called ``hyperon puzzle" of neutron stars~\cite{Vidana:2013nxa, Tolos:2020aln}.
However, such studies are very scarce due to the short life time of the hyperons. 
Recently, BESIII developed a novel method to study the hyperon-nucleon interactions, 
in which the hyperon–antihyperon pairs produced in charmonium decays serve as the source of hyperons, while
the detector materials themselves act as fixed targets~\cite{Dai:2024myk}.

This proceeding presents recent achievements in the aforementioned studies of hyperon physics at the BESIII experiment.

%
%
\section{Search for CP Violation in Hyperon Decays}

The hyperon non-leptonic weak decays provides a sensitive laboratory for testing CP symmetry. 
When a spin-1/2 hyperon undergoes weak decay into another spin-1/2 baryon and a pion, 
the resulting final state comprises both S-wave (parity-violating) and P-wave (parity-conserving) components. 
The decay amplitude can be formulated as $A = S + P\,\vec{\sigma}\cdot \hat{n}$, 
where $\hat{n}$ denotes the unit vector in the direction of the emitted baryon in the hyperon’s rest frame.

The dynamics of such weak decays can be described by decay parameters~\cite{Lee:1957qs}:
\begin{equation}
    \alpha = \frac{2\,\text{Re}(S^*P)}{|S|^2 + |P|^2}, \qquad \beta = \frac{2\,\text{Im}(S^*P)}{|S|^2 + |P|^2} = \sqrt{1-\alpha^2}\,\sin\phi,
\end{equation}
where $\alpha$, $\beta$, and $\phi$ encapsulate the CP-odd characteristics of the decay. 

If CP symmetry holds, these decay parameters for hyperons and antihyperons are expected to be equal in magnitude but opposite in sign, 
i.e., $\alpha = -\bar{\alpha}$, $\beta = -\bar{\beta}$, and $\phi = -\bar{\phi}$. 
To probe possible CP violation, the following observables can be constructed:
\begin{equation}
    A_{CP} = \frac{\alpha + \bar{\alpha}}{\alpha - \bar{\alpha}},\qquad \Delta\phi_{CP} = \frac{\phi + \bar{\phi}}{2},\qquad B_{CP} = \frac{\beta + \bar{\beta}}{\alpha - \bar{\alpha}} \approx (\xi_P-\xi_S). 
\end{equation}
Here, $(\xi_P-\xi_S)$ denotes the weak phase difference associated with CP violation. 
Observation of non-zero values for these quantities would constitute direct evidence for CP violation in hyperon decays.

\subsection{CP Violation in $\Lambda$ Decays}

The $\Lambda$ hyperon has long been a prime candidate for CP violation studies, owing to its straightforward decay channels and well-established properties. At BESIII, searches for CP symmetry breaking in $\Lambda$ decays are performed via the process $e^+e^- \to J/\psi \to \Lambda(\to p\pi^-) \bar{\Lambda} (\to \bar{p} \pi^+)$~\cite{BESIII:2022qax}. By reconstructing these events, a dataset containing 3.2 million quantum-entangled $\Lambda \bar{\Lambda}$ pairs has been assembled.
The joint angular distribution for this decay can be formulated as
\begin{equation}
    \begin{aligned}
        \mathcal{W}(\xi) = \mathcal{F}_{0}(\xi) + \alpha_{\psi} \mathcal{F}_{5}(\xi) + \alpha_{\Lambda} \bar{\alpha}_{\Lambda}\Big[ \mathcal{F}_{1}(\xi) + \sqrt{1-\alpha_{\psi}^{2}} \cos(\Delta \Phi)\, \mathcal{F}_{2}(\xi) + \alpha_{\psi} \mathcal{F}_{6}(\xi) \Big] \\
        + \sqrt{1-\alpha_{\psi}^{2}} \sin(\Delta \Phi) \left[ \alpha_{\Lambda} \mathcal{F}_{3}(\xi) + \bar{\alpha}_{\Lambda} \mathcal{F}_{4}(\xi) \right],
    \end{aligned}
    \label{eq:amplitude}
\end{equation}
where the functions $\mathcal{F}_i(\xi)$ $(i=0,1,...,6)$ are detailed in Ref.~\cite{Faldt:2017kgy}, 
the parameters $\alpha_{\psi}$ and $\Delta \Phi$ are related to the psionic form factors~\cite{Faldt:2017kgy} 
and govern the scattering angle distribution and the polarization of the $\Lambda$ and $\bar{\Lambda}$.

Using this formalism, a multi-dimensional angular analysis has been carried out, leveraging a five-dimensional likelihood fit to extract the relevant decay parameters, as summarized in Table~\ref{tab:Xi}. Here, $\langle \alpha_{\Lambda} \rangle = (\alpha_{\Lambda}-\bar{\alpha}_{\Lambda})/2$ denotes the average of 
the magnitude of $\alpha_{\Lambda}$ and $\bar{\alpha}_{\Lambda}$.

The analysis yields a value for the $\Lambda$ CP asymmetry: $A^{\Lambda}_{CP} = (-2.5 \pm 4.6 \pm 1.2) \times 10^{-3}$, representing the most stringent test of CP invariance in hyperon decays to date. Moreover, a pronounced transverse polarization of the $\Lambda$ has been observed in $J/\psi$ decays. 

BESIII has also probed CP violation in $\Sigma^+$ decays, the details can be found at Refs.~\cite{BESIII:2025fre},

\begin{table}[htbp]
    \caption{ The obtained parameter values for $\Lambda$, $\Xi^-$, and $\Xi^0$ channels.} 
    \label{tab:Xi}
\scalebox{0.9}
{
    \centering
    \begin{tabular}{l c c c}
        \hline \hline 
        Parameter                          &$J/\psi \to \Lambda \bar{\Lambda}$~\cite{BESIII:2022qax} &$J/\psi \to \Xi^- \bar{\Xi}^+$~\cite{BESIII:2021ypr} & $J/\psi \to \Xi^0 \bar{\Xi}^0$~\cite{BESIII:2023drj}\\
        \hline
        $\alpha_{J/\psi}$                   & 0.4748 $\pm$ 0.0022 $\pm$ 0.0031 & $\phantom{-}0.586\pm 0.012 \pm0.010$             & ${\color{white}-}0.514  \pm 0.006 \pm  0.015$      \\
        $\Delta\Phi$(rad)                   & 0.7521 $\pm$ 0.0042 $\pm$ 0.0066 & $\phantom{-}1.213\pm 0.046 \pm 0.016$            & ${\color{white}-}1.168  \pm 0.019 \pm  0.018$      \\
        $\alpha_{\Xi}$                      & -                                & $-0.376\pm0.007\pm0.003$                         & $-0.3750 \pm 0.0034 \pm 0.0016$                    \\
        $\bar{\alpha}_{\Xi}$                & -                                & $\phantom{-}0.371\pm0.007\pm0.002$               & ${\color{white}-}0.3790 \pm 0.0034 \pm 0.0021$     \\
        $\phi_{\Xi}$(rad)                   & -                                & $\phantom{-}0.011\pm0.019\pm0.009$               & ${\color{white}-}0.0051 \pm 0.0096 \pm 0.0018$     \\
        $\bar{\phi}_{\Xi}$(rad)             & -                                & $-0.021\pm0.019\pm0.007$                         & $-0.0053 \pm 0.0097 \pm 0.0019$                    \\
        $\alpha_{\Lambda}$                  & 0.7519 $\pm$ 0.0036 $\pm$ 0.0024 & $\phantom{-}0.757 \pm 0.011 \pm 0.008$           & ${\color{white}-}0.7551 \pm 0.0052 \pm 0.0023$     \\
        $\bar{\alpha}_{\Lambda}$            & -0.7559 $\pm$ 0.0036 $\pm$ 0.0030& $-0.763 \pm 0.011 \pm 0.007$                     & $-0.7448 \pm 0.0052 \pm 0.0017$                    \\
        \hline                                   
        $\xi_{P}-\xi_{S}$(rad)              & -                                & $\phantom{-}(1.2\pm3.4\pm0.8)\times10^{-2}$      & ${\color{white}-}(0.0  \pm 1.7  \pm 0.2)\times 10^{-2}$ \\ 
        $\delta_{P}-\delta_{S}$(rad)        & -                                & $(-4.0\pm3.3\pm1.7)\times10^{-2}$                & $(-1.3  \pm 1.7  \pm 0.4)\times 10^{-2}$                \\ 
        \hline 
        $A^{\Xi}_{CP}$                      & -                                & $\phantom{-}(6.0\pm13.4\pm5.6)\times10^{-3}$     & $(-5.4 \pm 6.5 \pm 3.1)\times 10^{-3}$                  \\  
        $\Delta\phi^{\Xi}_{CP}$(rad)        & -                                & $(-4.8\pm13.7\pm2.9)\times10^{-3}$               & $(-0.1 \pm 6.9 \pm 0.9)\times 10^{-3}$    \\ 
        $A^{\Lambda}_{CP}$                  & -0.0025 $\pm$ 0.0046 $\pm$ 0.0012& $(-3.7\pm11.7\pm9.0)\times10^{-3}$               & ${\color{white}-}(6.9 \pm 5.8 \pm 1.8) \times 10^{-3}$  \\  
        \hline 
        $\langle \alpha_{\Xi} \rangle$      & -                                & -                                                & $-0.3770 \pm 0.0024\pm 0.0014$                 \\ 
        $\langle \phi_{\Xi} \rangle$(rad)   & -                                & $\phantom{-}0.016 \pm 0.014 \pm 0.007$           & ${\color{white}-}0.0052 \pm 0.0069 \pm 0.0016$ \\ 
        $\langle \alpha_{\Lambda} \rangle$  & 0.7542 $\pm$ 0.0010 $\pm$ 0.0024 & -                                                & ${\color{white}-}0.7499 \pm 0.0029 \pm 0.0013$ \\
        \hline \hline
    \end{tabular}
    }
\end{table}

\subsection{CP Violation in $\Xi$ Decays}
In contrast to $\Lambda$ and $\Sigma^+$ decays, the sequential decay chains of double-strange $\Xi$ hyperons are particularly compelling, as they enable the disentanglement of weak phase differences. At BESIII, comprehensive angular analyses have been conducted using large datasets: for the process $e^+ e^- \to J/\psi \to \Xi^-(\to \Lambda \pi^-) \bar{\Xi}^+ (\to \bar{\Lambda} \pi^+), \Lambda\to p\pi^-, \bar{\Lambda}\to \bar{p}\pi^+$, a sample based on 1.3 billion $J/\psi$ events was analyzed~\cite{BESIII:2021ypr}; for $e^+ e^- \to J/\psi \to \Xi^0(\to \Lambda \pi^0) \bar{\Xi}^0 (\to \bar{\Lambda} \pi^0), \Lambda\to p\pi^-, \bar{\Lambda}\to \bar{p}\pi^+$, 10 billion $J/\psi$ events were used~\cite{BESIII:2023drj}.

The multi-dimensional joint angular distribution, which encapsulates the spin correlations and polarization of the produced hyperons, is expressed as
\begin{equation}
    \label{eq:W}
    W(\xi; \omega)
    = \sum^{3}_{\mu, \bar{\nu}=0}{\rm C}_{\mu\bar{\nu}}(\theta_{\Xi}; \alpha_{\psi}, \Delta{\Phi})\sum^{3}_{\mu'=0}\sum^{3}_{\nu'=0}a^{\Xi}_{\mu\mu'}a^{\Lambda}_{\mu'0}a^{\bar{\Xi}}_{\nu\nu'}a^{\bar{\Lambda}}_{\nu'0},
\end{equation}
the details of the formula can be found in Ref.~\cite{Perotti:2018wxm}.


Utilizing this formalism, nine-dimensional likelihood fits were performed to extract physical parameters for both $\Xi^-$ and $\Xi^0$ channels. The results are summarized in Table~\ref{tab:Xi}. No evidence for CP violation was observed within the sensitivity of $10^{-3}$, establishing stringent constraints and providing an essential benchmark for future new physics searches in hyperon weak decays. Furthermore, BESIII has, for the first time, observed significant polarization for both $\Xi^0$ and $\Xi^-$ produced in $J/\psi$ decays.

\subsection{CP Violation in $\Sigma^0$ Decays}

The $\Sigma^0$ hyperon holds a distinctive position among the ground-state hyperons, 
as it is the only one that decays exclusively via radiative processes. 
Notably, the electric dipole transition moment in its decay $\Sigma^0 \to \Lambda \gamma$ is linked to the neutron electric dipole moment (EDM) 
through SU(3) flavor symmetry considerations~\cite{Gell-Mann:1961omu, Nair:2018mwa}. 
As a result, studies of the radiative decay of the $\Sigma^0$ can provide valuable constraints on possible sources of strong CP violation~\cite{Nair:2018mwa}.

At BESIII, a detailed angular analysis has been performed using $e^+ e^- \to \psi \to \Sigma^0(\to \Lambda \gamma) \bar{\Sigma}^0 (\to \bar{\Lambda} \gamma)$,
followed by $\Lambda \to p\pi^-$ and $\bar{\Lambda} \to \bar{p}\pi^+$. 
This study utilizes data samples of 10 billion $J/\psi$ and 2.7 billion $\psi(3686)$ decays, 
where $\psi$ denotes either $J/\psi$ or $\psi(3686)$~\cite{BESIII:2024nif}. 
The joint angular distribution captures the spin correlations and polarization features of the process, and is given by
\begin{align}
    \label{formulation}
    & \mathcal{W}(\vec{\zeta}, \vec{\omega})
    \propto (1-\alpha_{\Lambda}\alpha_{\Sigma^0}\cos\theta_p)(1-\bar{\alpha}_{\Lambda}\bar{\alpha}_{\Sigma^0}\cos\theta_{\bar{p}})\times \notag \\
    & \{1 + \alpha_{\psi}\cos^2\theta_{\Sigma} \notag + \sqrt{1-\alpha^2_{\psi}}\sin\theta_{\Sigma} \cos\theta_{\Sigma}  
    \cdot \notag \\
    & [\beta_{\gamma}\bar{\beta}_{\gamma} F_1 - (\beta_{\gamma} F_2 - \bar{\beta}_{\gamma} F_3)] + \notag \\
    & \beta_{\gamma}\bar{\beta}_{\gamma} [\alpha_{\psi} F_4 + F_5 - 
    (\alpha_{\psi} + \cos^2 \theta_{\Sigma})\cos\theta_{\Lambda}\cos\theta_{\bar{\Lambda}}]\},
\end{align}
the details can be found in Ref.~\cite{BESIII:2024nif}.

According Eq.~\ref{formulation}, a detailed full-angular analysis has been conducted, 
and a seven-dimensional likelihood fit has been performed to determine the parameter values. 
The strong CP asymmetry in $\Sigma^0$ decay has been determined to be $A^{\Sigma^0}_{CP} = \alpha_{\Sigma^0} + \bar{\alpha}_{\Sigma^0}
= (0.4 \pm 2.9 \pm 1.3) \times 10^{-3}$, which remarks the first test of strong CP symmetry in the hyperon decays. 
Furthermore, the transverse polarizations of the $\Sigma^0$ have been observed in $J/\psi$ and $\psi(3686)$ decays with opposite directions.

\section{Hyperon Weak Radiative and Semi-leptonic Decays}
\begin{figure}[t]
    \centering
    \includegraphics[width=5.2cm]{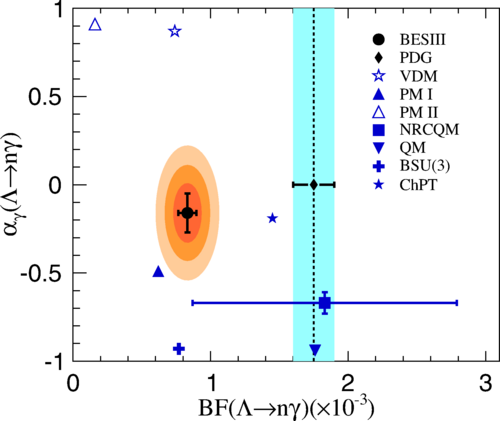}
    \includegraphics[width=6cm]{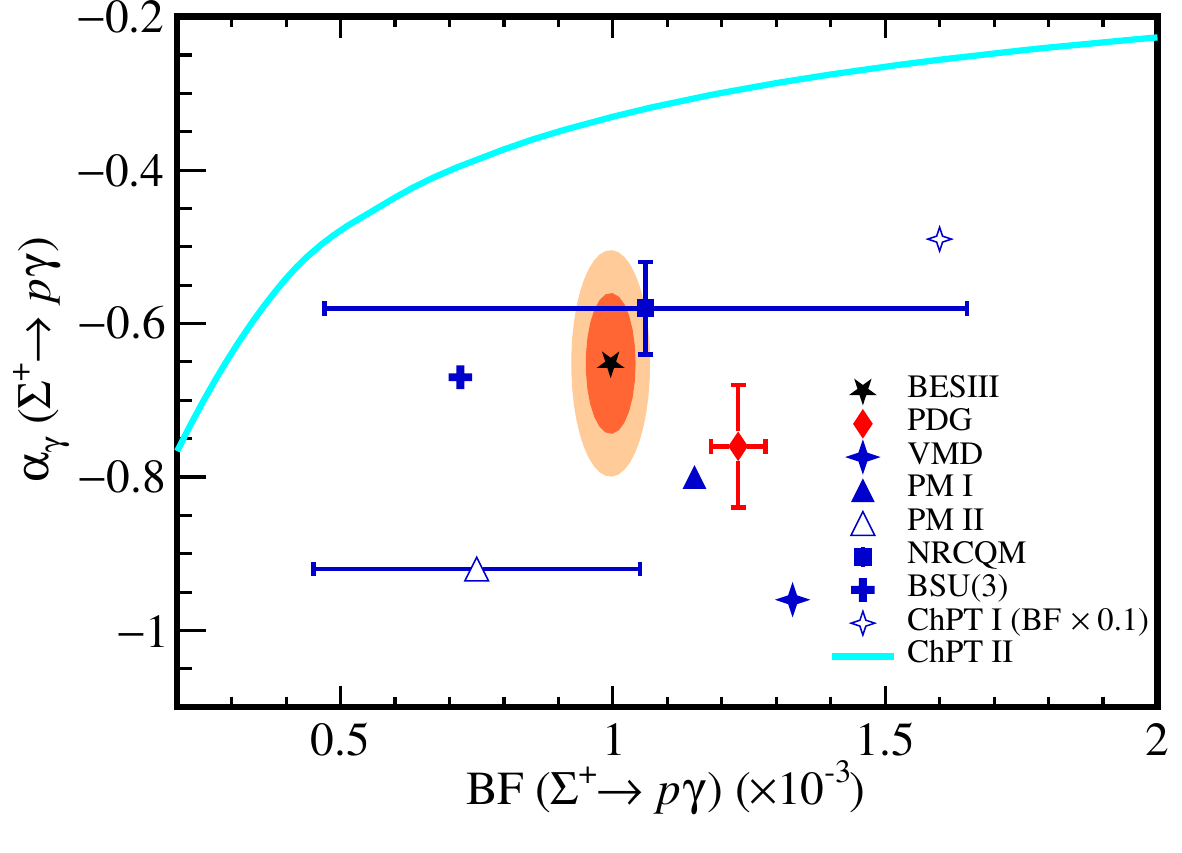}
    \caption{Two dimensional distributions of the BFs and decay parameters of $\Lambda\to n\gamma$(left) and $\Sigma^+\to p\gamma$(right) decays.}
    \label{f1}
\end{figure}

The hyperon weak radiative decays $\Lambda\to n\gamma$~\cite{BESIII:2022rgl} and $\Sigma^+\to p\gamma$~\cite{BESIII:2023fhs} have been studied at BESIII
in the decays $J/\psi \to \Lambda\bar{\Lambda}$ and $J/\psi \to \Sigma^+ \bar{\Sigma}^-$. 
The absolute branching fractions (BFs) of these two decays are measured for the first time, as well as their decay parameters. 
Figure~\ref{f1} shows the results of BFs and decay parameters for $\Lambda\to n\gamma$ and $\Sigma^+\to p\gamma$, respectively, 
including the previous PDG results and theoretical predictions.
Of note, our results $BF_{\Lambda \to n \gamma} = (0.832 \pm 0.038 \pm 0.054) \times 10^{-3}$ and 
$BF_{\Sigma^+ \to p \gamma} = (0.996 \pm 0.021 \pm 0.018) \times 10^{-3}$ both lower than the previous PDG values by $5.6\sigma$ and $4.2\sigma$, respectively.

The absolute BF of the semi-leptonic decay $\Lambda \to p\mu^- \bar{\nu}_{\mu}$ has been measured for the first time at BESIII~\cite{BESIII:2021ynj}, 
which is $BF_{\Lambda \to p\mu^- \bar{\nu}_{\mu}} = (1.48 \pm 0.21 \pm 0.08) \times 10^{-4}$. 
Combining the PDG value of the BF of $\Lambda \to p e^- \bar{\nu}_{e}$ ($BF_{\Lambda \to p e^- \bar{\nu}_{e}}$), 
a test of lepton flavor universality has been conducted, 
$R^{\mu e}=BF_{\Lambda \to p\mu^- \bar{\nu}_{\mu}}/BF_{\Lambda \to p e^- \bar{\nu}_{e}} = 0.178\pm0.028$,
which is consistent with the conservation of lepton flavor universality.

\section{Hyperon-nucleon Interaction}
\begin{figure}[t]
    \centering
    \includegraphics[width=6cm]{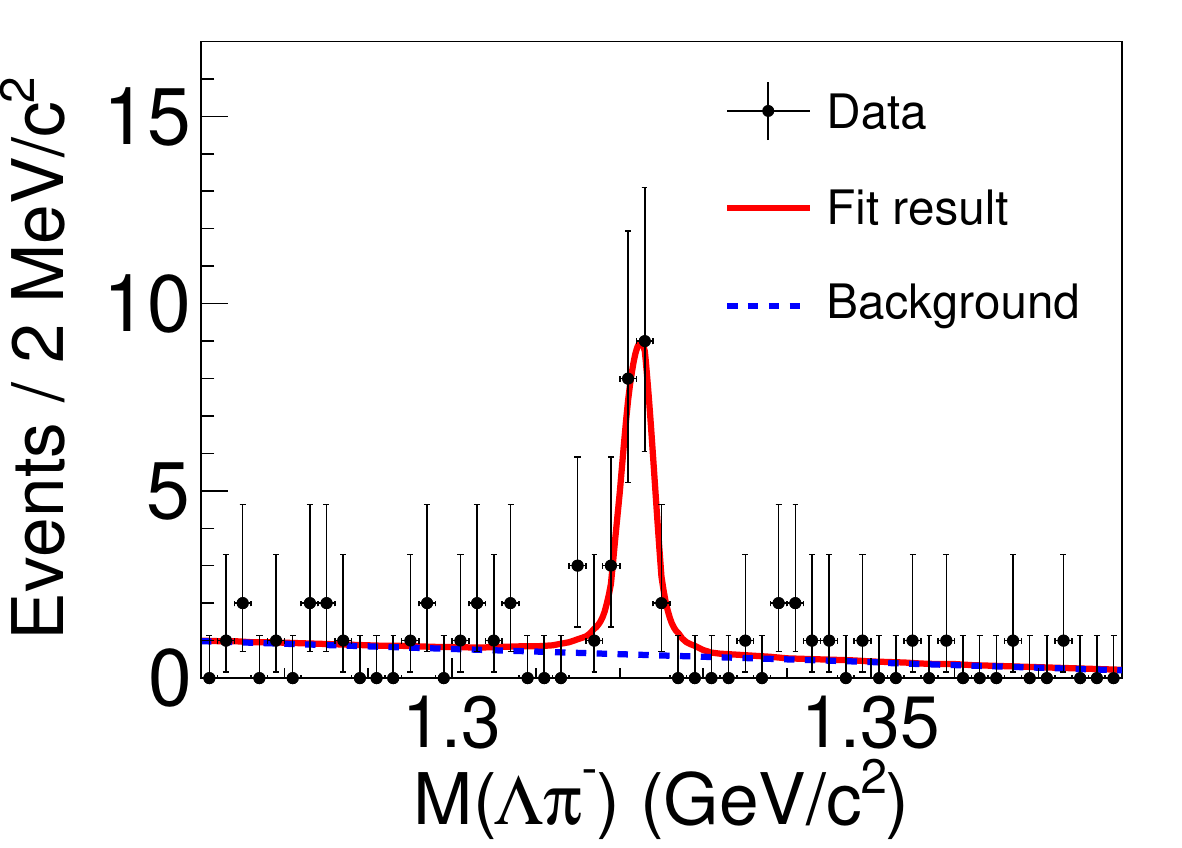}
    \caption{Distribution of the $M_{\Lambda \pi^-}$.}
    \label{f2}
\end{figure}

In 2023, BESIII reported the first study of reaction $\Xi^0 n \to \Xi^- p$ using $\Xi^0$-nucleus scattering at an electron-positron collider, 
and opened up a new direction for such researches~\cite{BESIII:2023clq}.
The $\Xi^0$ beam is produced from the $J/\psi \to \Xi^0 \bar{\Xi}^0$, then the $\Xi^0$ hyperons interact with the materials in the beam pipe.
Figure~\ref{f2} shows the distribution of signal yields of the $\Xi^-$ produced by $\Xi^0 + n$ interaction.
The cross section for the reaction $\Xi^0 + {}^{9}\mathrm{Be} \to \Xi^- + p + {}^{8}\mathrm{Be}$ has been measured to be $22.1 \pm 5.3 \pm 4.5$~mb at a $\Xi^0$ momentum of $0.818$ GeV/$c$.
In addition,
the cross sections of $\Lambda + {}^9\mathrm{Be} \to \Sigma^+ + X$, $\Lambda + p\to \Lambda + p$, and $\bar{\Lambda} + p \to \bar{\Lambda} p$ 
have also been measured at BESIII~\cite{BESIII:2023trh, BESIII:2024geh}.

\section{Search for Hyperon EDM in $J/\psi$ Decays}

The EDM of elementary particles, which arises from flavor-diagonal $CP$ violation, 
is a sensitive probe for physics beyond the Standard Model and may offer crucial insights into the matter-antimatter asymmetry of the universe. 
However, due to the short lifetimes, hyperon EDM remains a largely unexplored territory. 
The entangled nature of the hyperon-antihyperon pairs provides a unique opportunity to extract hyperon EDMs indirectly~\cite{He:2022jjc, Fu:2023ose}.

Utilizing this approach, BESIII has conducted a search for the $\Lambda$ EDM via the process $J/\psi \to \Lambda(\to p\pi^-)\, \bar{\Lambda} (\to \bar{p} \pi^+)$~\cite{BESIII:2025vxm}. 
The upper limit is established at the 95\% confidence level:
\begin{equation}
    \begin{aligned}
        |d_{\Lambda}| < 6.5\times 10^{-19}~e~\text{cm}.
    \end{aligned}
\end{equation}
This result for the $\Lambda$ EDM improves upon the previous limit set by Fermilab in 1981~\cite{Pondrom:1981gu} by three orders of magnitude, 
placing stringent constraints on new physics scenarios involving CP violation that aim to explain the dominance of matter in the universe.

\section{Summary}

Since its inception, 
BESIII has been operating successfully and has accumulated substantial data samples in the $\tau$-charm physics region. 
Significant progress has been achieved in hyperon physics, including precise CP tests in the two-body weak decays of $\Lambda$, $\Sigma$, and $\Xi$ hyperons, 
precise measurements of absolute BFs of the hyperon weak radiative and semi-leptonic decays, studies of hyperon-nucleon interactions, 
and measurement of the $\Lambda$ EDM.
In the future, BESIII will continue to deliver high-quality researches to the physics community.

\section*{Acknowledgement}
This work is supported in part by National Key R\&D Program of China under Contracts No. 2023YFA1606000; 
National Natural Science Foundation of China (NSFC) under Contracts Nos. 11935018, 12342502; 
China Postdoctoral Science Foundation under Grant No. 2024M753040; 
Postdoctoral Fellowship Program of China Postdoctoral Science Foundation under Grant No. GZC20241608;
Beijing Natural Science Foundation of China (BNSF) under Contract No. IS23014.

\end{document}